# Accelerated Cardiac Diffusion Tensor Imaging Using Joint Low-Rank and Sparsity Constraints

Sen Ma, Christopher T. Nguyen, Anthony G. Christodoulou, *Member, IEEE*, Daniel Luthringer, Jon Kobashigawa, Sang-Eun Lee, Hyuk-Jae Chang and Debiao Li

*Abstract— Objective:* The purpose of this manuscript is to accelerate cardiac diffusion tensor imaging (CDTI) by integrating low-rankness and compressed sensing. *Methods:* Diffusion-weighted images exhibit both transform sparsity and low-rankness. These properties can jointly be exploited to accelerate CDTI, especially when a phase map is applied to correct for the phase inconsistency across diffusion directions, thereby enhancing low-rankness. The proposed method is evaluated both ex vivo and in vivo, and is compared to methods using either a low-rank or sparsity constraint alone. *Results:* Compared to using a low-rank or sparsity constraint alone, the proposed method preserves more accurate helix angle features, the transmural continuum across the myocardium wall, and mean diffusivity at higher acceleration, while yielding significantly lower bias and higher intraclass correlation coefficient. *Conclusion:* Low-rankness and compressed sensing together facilitate acceleration for both ex vivo and in vivo CDTI, improving reconstruction accuracy compared to employing either constraint alone. *Significance:* Compared to previous methods for accelerating CDTI, the proposed method has the potential to reach higher acceleration while preserving myofiber architecture features which may allow more spatial coverage, higher spatial resolution and shorter temporal footprint in the future.

## I. INTRODUCTION

Cardiac diffusion tensor imaging (CDTI) is a powerful noninvasive tool capable of assessing the anatomic microstructure of the myocardium which is highly structured and organized into sheets of fibers making it suitable to be characterized by CDTI [1-3]. CDTI, performed both ex vivo [4-6] and in vivo [7, 8], reveals a helical fiber pattern along the ventricle wall with left-handed orientation in the subepicardial region and right-handed orientation in the subendocardial region for healthy heart [9]. Such pattern can be characterized by helix angle (HA) which represents the elevated angle out of the short-axis plane, indicating the local fiber orientation. Myofibers around subendocardial regions, mid myocardium and subepicardial regions have HA$> 0°$, HA$= 0°$ and HA$< 0°$, respectively [2]. In heart failure, the helical structure and orientation of the myocardial fibers are severely altered due to adverse remodeling [10, 11].

One of the major challenges for CDTI is the prolonged acquisition time, because of the multiple diffusion encoding measurements needed to robustly reconstruct the self-diffusion tensor [12]. In addition, multiple signal averages are required to maintain sufficient signal-to-noise ratio (SNR) due to signal loss caused by $T_2$ decay and diffusion signal attenuation [13]. Though motion-induced signal loss can be effectively addressed by second- or higher-order motion compensation gradient waveforms [7, 14], long acquisitions will likely incur more complex motion and patient discomfort precluding robust clinical translation of CDTI.

One approach for accelerated data acquisition is sparse modeling based on compressed sensing (CS) [15, 16], which enables recovery of missing data of highly undersampled k-space measurements using nonlinear reconstruction. This works by exploiting signal sparsity in a transform domain in which undersampling artifacts are incoherent [17, 18]. Compressed sensing approaches have a wide range of applications [19-21] and have been used in CDTI to provide precise measurements of fractional anisotropy (FA), mean diffusivity (MD), and the primary eigenvector ($\Delta\alpha$) until 4× acceleration [22].

Another approach is low-rank modeling (LR) which exploits signal correlation using partial separability model [23, 24]. It has previously been used for diffusion-weighed image denoising in the brain [25] as well as dynamic cardiac imaging at high acceleration factors [26]. Additionally, Gao *et al.* proposed a phase-constrained low-rank method to accelerate

Manuscript received July 25, 2017; revised November 21, 2017; accepted December 18, 2017. This work was supported by the U.S. National Institutes of Health under Grants 1R01HL124649, R21 EB024701-01 and T32HL116273.
S. Ma is with the Department of Bioengineering, University of California, Los Angeles, USA, and also with the Biomedical Imaging Research Institute, Cedars-Sinai Medical Center, USA. C. T. Nguyen is with the Cardiovascular Research Center, Massachusetts General Hospital, USA, and also with the Biomedical Imaging Research Institute, Cedars-Sinai Medical Center, USA. A. G. Christodoulou is with the Biomedical Imaging Research Institute, Cedars-Sinai Medical Center, USA. D. Luthringer is with the Department of Pathology, Cedars-Sinai Medical Center, USA. J. Kobashigawa is with the Cedars-Sinai Heart Institute, Cedars-Sinai Medical Center, USA. S. E. Lee is with the Biomedical Imaging Research Institute, Cedars-Sinai Medical Center, USA, Severance Cardiovascular Hospital, Republic of Korea, Yonsei-Cedars-Sinai Integrative Cardiovascular Imaging Research Center, Republic of Korea, and Yonsei University College of Medicine, Republic of Korea. H. J. Chang is with Severance Cardiovascular Hospital, Republic of Korea, Yonsei-Cedars-Sinai Integrative Cardiovascular Imaging Research Center, Republic of Korea, and Yonsei University College of Medicine, Republic of Korea. D. Li is with the Biomedical Imaging Research Institute, Cedars-Sinai Medical Center, USA, and also with the Department of Bioengineering, University of California, Los Angeles, USA. (e-mail: debiao.li@cshs.org).

brain DTI, which lowered the root-mean-squared error (RMSE) of diffusion-weighted images, FA and MD compared to FFT reconstruction at 4× acceleration [27]. However, the phase correction component of this method requires dense sampling at the center of k-space, limiting the potential for further acceleration.

Because diffusion-weighted iamges exhibit both low-rank structure and transform sparsity, LR and CS approaches can be integrated to express the images using even fewer degrees of freedom, offering a higher potential for acceleration. Various forms of sparse and low-rank combinations have shown promising results when accelerating dynamic MRI, parameter mapping and 4D flow MRI [28-32]. Huang *et al.* combined an implicit low-rank constraint and a joint sparsity constraint to accelerate CDTI which, for real human heart data, reduced the RMSE of FA and MD at 5× acceleration compared to using basic compressed sensing, joint sparsity constraint alone and low-rank constraint alone [33]. However, the authors did not conduct evaluation on HA, nor performed a phase correction step to compensate for the drastic eddy current-induced phase inconsistency between diffusion directions that reduces correlation and weakens low-rankness (as described in [27]).

In this work, we present a phase-corrected LR and CS approach for accelerating CDTI. Specifically, we incorporate an explicit-subspace low-rank component and a group sparsity component into a unified framework. Furthermore, we estimate a phase map for correction using the full undersampled data, as opposed to from low-resolution scans as was done in [27]. We show that the joint combination of LR and CS provides higher potential for acceleration, better image quality, and higher reconstruction accuracy. We test our method ex vivo on six human heart failure cadavers and in vivo in seven hypertrophic cardiomyopathy patients, measuring HA and MD to evaluate the proposed method's ability to preserve features present in heart disease, its ability to maintain helical fiber structure and the transmural change from endocardium to epicardium, and its potential to reduce scan time.

## II. THEORY

### A. Image Model

*1) Low-rank constraint with phase correction:* CDTI is performed by acquiring a sequence of complex-valued diffusion-weighted images $\{x(\mathbf{r},d)\}_{d=1}^{N}$ with $N$ diffusion encoding directions and/or diffusion weightings. Due to the strongly correlated behaviors of diffusion-weighted signals at different voxels (similar to some dynamic MRI scenarios [23, 24]), the diffusion-weighted images can be modeled as $L$th-order partially-separable:

$$x(\mathbf{r},d) = \sum_{\ell=1}^{L} u_\ell(\mathbf{r}) v_\ell(d) \quad (1)$$

where $\{v_\ell(\cdot)\}_{\ell=1}^{L}$ are the diffusion basis functions and $\{u_\ell(\cdot)\}_{\ell=1}^{L}$ are the corresponding spatial coefficients. If we rearrange the elements of the diffusion-weighted image sequence $\{x(\mathbf{r},d)\}_{d=1}^{N}$ as a matrix where the rows and columns represent the spatial and diffusion dimension respectively, the resulting Casorati matrix [23] is:

$$\mathbf{X} = \begin{bmatrix} x(\mathbf{r}_1, 1) & \cdots & x(\mathbf{r}_1, N) \\ \vdots & \ddots & \vdots \\ x(\mathbf{r}_M, 1) & \cdots & x(\mathbf{r}_M, N) \end{bmatrix} \in \mathbb{C}^{M \times N} \quad (2)$$

where $M$ denotes the total number of voxels. Based on (1), $\mathbf{X}$ is low-rank when $L < \min\{M, N\}$. This low-rank structure can be explicitly expressed through matrix factorization as $\mathbf{X} = \mathbf{UV}$, where $\mathbf{V} \in \{\mathbb{C}^{L \times N}: V_{jk} = v_j(k)\}$ contains "diffusion basis functions" spanning the low-dimensional diffusion subspace and $\mathbf{U} \in \{\mathbb{C}^{M \times L}: U_{jk} = u_k(\mathbf{r}_j)\}$ contains the corresponding spatial coefficients spanning the spatial subspace. This low-rank structure offers considerable potential for undersampling because $\mathbf{X}$ has $MN$ complex elements ($2MN$ real values) but only $2(M + N - L)L$ degrees of freedom. Image reconstruction can be performed by finding a low-rank matrix $\mathbf{X}$ which is consistent with the data [23], or by estimating $\mathbf{U}$ and $\mathbf{V}$ in separate steps [24]. In the latter strategy, which has not yet been applied to CDTI, $\mathbf{V}$ is typically estimated first, forming a subspace constraint which explicitly enforces low-rankness.

In practice, the low-rank property of $\mathbf{X}$ may be weakened by uncorrelated phase changes for different diffusion encoding directions. Low-rankness can be restored by modeling these phase inconsistencies in the form of a unit-magnitude phase map $\mathbf{P} \in \{\mathbb{C}^{M \times N}: |P_{jk}| = 1, \forall j, k\}$ which in previous work has been calculated from low-resolution scans [27]. The phase-corrected image model is thus $\mathbf{X} = \mathbf{P} \circ (\mathbf{UV})$, where $\circ$ denotes Hadamard (elementwise) multiplication.

*2) Group sparsity constraint:* Group sparsity modeling is inspired by distributed compressed sensing and has been applied on its own to accelerate CDTI [22]. The underlying assumption is that an individual image in the diffusion-weighted image series not only has a sparse property in some transform domain calculated by applying the matrix $\mathbf{\Psi}$, but also shares similar sparse support with other images. In other words, the sparse coefficients of consecutive diffusion-weighted images in the transform domain are correlated in the diffusion encoding dimension. This can be accomplished by solving a problem of the form

$$\arg\min_{\mathbf{X}} \|\mathbf{d} - E(\mathbf{X})\|_2^2 + \lambda R_s(\mathbf{X}) \quad (3)$$

where $\mathbf{d}$ is a vector of undersampled k-space data, $E(\cdot)$ performs spatial encoding and sparse sampling, and $R_s(\cdot)$ is the regularization penalty promoting group sparsity. Group sparsity is characterized by an $\ell_{1,2}$-norm penalty, i.e., $R_s(\mathbf{X}) = \|\mathbf{\Psi X}\|_{1,2}$, where $\|\mathbf{Y}\|_{1,2} = \sum_{i=1}^{n} \|\mathbf{Y}^{(i)}\|_2$ and $\mathbf{Y}^{(i)}$ is the $i$th group in $\mathbf{Y}$. For example, the sparse coefficients at one voxel along the diffusion encoding dimension can be considered as one group, which gives an explicit expression of

$$R_s(\mathbf{X}) = \sum_{j=1}^{M} \sqrt{\sum_{k=1}^{N} |[\mathbf{\Psi X}]_{jk}|^2}. \quad (4)$$

*3) Relationship between constraints:* It is worth mentioning that the low-rank constraint and group sparsity constraint are complementary. At high acceleration factors, the low-rank constraint alone can result in an ill-conditioned problem, causing severe image artifacts; the group sparsity constraint alone can cause image blurring and induce inaccurate measurements. However, the group sparsity constraint can be used to regularize the low-rank problem, reducing image artifacts while facilitating undersampling. The low-rank constraint takes advantage of the signal correlation while the group sparsity constraint promotes sparse solutions, thus improving reconstruction accuracy.

### B. Problem Formulation

Here we propose a mathematical framework that reconstructs the cardiac diffusion-weighted image sequence from highly undersampled k-space data using joint low-rank and sparsity constraints and a phase correction procedure. We express the signal equation from $C$ receiver coils as:

$$\mathbf{d} = \Omega(\mathbf{FS}[\mathbf{P} \circ (\mathbf{UV})]) + \boldsymbol{\eta} \quad (5)$$

where $\mathbf{d} \in \mathbb{C}^{D \times 1}$ is the vector of undersampled multi-coil k-space data comprising $D < MNC$ total samples, $\Omega(\cdot): \mathbb{C}^{MC \times N} \to \mathbb{C}^{D \times 1}$ denotes the k-space undersampling operator, $\mathbf{S} \in \mathbb{C}^{MC \times M}$ applies the coil sensitivity maps to produce individual coil images, $\mathbf{F} \in \mathbb{C}^{MC \times MC}$ is a block-diagonal operator which applies the spatial Fourier transform to each coil image independently and $\boldsymbol{\eta} \in \mathbb{C}^{D \times 1}$ represents measurement noise. With this signal model, the joint low-rank and group sparsity-constrained reconstruction problem can be formulated as

$$\arg\min_{\mathbf{P},\mathbf{U},\mathbf{V}} \|\mathbf{d} - \Omega(\mathbf{FS}[\mathbf{P} \circ (\mathbf{UV})])\|_2^2 + \lambda \|\mathbf{\Psi UV}\|_{1,2} \quad (6)$$

where the resulting diffusion-weighted image sequence is given as $\mathbf{X} = \mathbf{UV}$. We propose to estimate $\mathbf{P}$, $\mathbf{V}$ and $\mathbf{U}$ in three separate steps.

*1) Estimate phase map:* Firstly, we propose to estimate the phase map $\widehat{\mathbf{P}}$ from a preliminary reconstruction by enforcing only the group sparsity constraint, i.e.,

$$\widetilde{\mathbf{X}} = \arg\min_{\mathbf{X}} \|\mathbf{d} - \Omega(\mathbf{FSX})\|_2^2 + \lambda R_s(\mathbf{X}) \quad (7)$$

which allows construction of $\widehat{\mathbf{P}}$ from the phase of $\widetilde{\mathbf{X}}$ according to $\widehat{P}_{jk} = \exp(i\angle \widetilde{X}_{jk})$.

*2) Estimate diffusion subspace:* Secondly, we construct a matrix $\widehat{\mathbf{V}}$ from the singular value decomposition (SVD) of the magnitude image $|\widetilde{\mathbf{X}}|$, by collecting and transposing the $L$ most significant right singular vectors. This leaves only the spatial coefficient matrix to be recovered.

*3) Recover spatial coefficient matrix:* Lastly, we recover the spatial coefficient matrix according to the updated (6):

$$\widehat{\mathbf{U}} = \arg\min_{\mathbf{U}} \|\mathbf{d} - \Omega(\mathbf{FS}[\widehat{\mathbf{P}} \circ (\mathbf{U}\widehat{\mathbf{V}})])\|_2^2 + \lambda \|\mathbf{\Psi U}\widehat{\mathbf{V}}\|_{1,2}. \quad (8)$$

We will demonstrate the advantage of incorporating a low-rank component to the compressed sensing framework (LR/CS) over using compressed sensing framework (CS Only) or low-rank modeling (LR Only) alone in the Results section.

### C. Algorithm

Note that (8) reduces to (7) if $\widehat{\mathbf{P}}$ is set to be identity matrix and $\widehat{\mathbf{V}}$ is full rank (i.e., $L = N$ or rank($\widehat{\mathbf{V}}$) = $N$). Without loss of generality, this subsection will just provide procedures of solving (8) which is a non-smooth convex optimization problem and can be solved by a variety of algorithms [34-36]. Here we adopt the alternating direction method of multipliers (ADMM) algorithm which is an efficient and fast algorithm well suited for solving large-scale optimization problems [37].

*1) Summary of algorithm:* We derive an equivalent problem for (8) by change of variables:

$$\min_{\mathbf{U}} \|\mathbf{d} - \Omega(\mathbf{FS}[\widehat{\mathbf{P}} \circ (\mathbf{U}\widehat{\mathbf{V}})])\|_2^2 + \lambda \|\mathbf{G}\|_{1,2} \quad (9)$$
$$\text{s.t. } \mathbf{\Psi U}\widehat{\mathbf{V}} - \mathbf{G} = 0.$$

The augmented Lagrangian function for (9) can be written as:

$$L(\mathbf{U}, \mathbf{G}, \mathbf{Y}) = \|\mathbf{d} - \Omega(\mathbf{FS}[\widehat{\mathbf{P}} \circ (\mathbf{U}\widehat{\mathbf{V}})])\|_2^2 + \lambda \|\mathbf{G}\|_{1,2} \\ + \frac{\rho}{2} \|\mathbf{\Psi U}\widehat{\mathbf{V}} - \mathbf{G}\|_2^2 + \mathbf{Y}^H(\mathbf{\Psi U}\widehat{\mathbf{V}} - \mathbf{G}), \quad (10)$$

where $\mathbf{Y}$ is the Lagrangian multiplier and $\rho$ is a regularization parameter controlling the speed of convergence. Equation (10) can be minimized through alternating iterations: assuming $\mathbf{U}^k$, $\mathbf{G}^k$ and $\mathbf{Y}^k$ are fixed solutions from the $k$th iteration, then for $(k + 1)$th iteration we need to solve:

$$\mathbf{G}^{k+1} = \underset{\mathbf{G}}{\text{argmin}}\, L(\mathbf{U}^k, \mathbf{G}, \mathbf{Y}^k), \quad (11)$$
$$\mathbf{U}^{k+1} = \underset{\mathbf{U}}{\text{argmin}}\, L(\mathbf{U}, \mathbf{G}^{k+1}, \mathbf{Y}^k), \quad (12)$$
$$\mathbf{Y}^{k+1} = \mathbf{Y}^k + \rho(\mathbf{\Psi U}^{k+1}\widehat{\mathbf{V}} - \mathbf{G}^{k+1}). \quad (13)$$

*2) Solution of (11):* Sub-problem (11) can be rewritten as:

$$\mathbf{G}^{k+1} = \underset{\mathbf{G}}{\text{argmin}}\, \lambda\|\mathbf{G}\|_{1,2} + \frac{\rho}{2}\|\mathbf{Z}^k - \mathbf{G}\|_2^2, \quad (14)$$

where $\mathbf{Z}^k = \mathbf{\Psi U}^k\widehat{\mathbf{V}} + \mathbf{Y}^k/\rho$. Equation (14) is separable with respect to groups of $\mathbf{G}$ which is equivalent to:

$$\mathbf{G}_{(i)}^{k+1} = \underset{\mathbf{G}_{(i)}}{\text{argmin}}\, \lambda\|\mathbf{G}_{(i)}\|_2 + \frac{\rho}{2}\|\mathbf{Z}_{(i)}^k - \mathbf{G}_{(i)}\|_2^2, \quad (15)$$
$$i = 1,2,\dots,n$$

where $\mathbf{G}_{(i)}$ and $\mathbf{Z}_{(i)}^k$ denotes the $i$th group of $\mathbf{G}$ and $\mathbf{Z}^k$ and $n$ is the number of groups in $\mathbf{G}$. The solution to (15) is given by:

$$\mathbf{G}_{(i)}^{k+1} = \frac{\mathbf{Z}_{(i)}^k}{\|\mathbf{Z}_{(i)}^k\|_2} \mathcal{S}_\alpha\left(\|\mathbf{Z}_{(i)}^k\|_2\right), \quad (16)$$

where $\alpha = \lambda/\rho$, and $\mathcal{S}_\alpha(\cdot)$ is the soft-thresholding operator defined as:

$$\mathcal{S}_\alpha(x) = \begin{cases} x - \alpha & \text{if } x > \alpha \\ 0 & \text{if } |x| \leq \alpha. \\ x + \alpha & \text{if } x < \alpha \end{cases} \quad (17)$$

*3) Solution of (12):* Sub-problem (12) can be rewritten as:

$$\mathbf{U}^{k+1} = \underset{\mathbf{U}}{\text{argmin}} \|\mathbf{d} - \mathcal{A}(\mathbf{U})\|_2^2 + \frac{\rho}{2}\|\mathcal{B}(\mathbf{U}) - \mathbf{G}^{k+1}\|_2^2 \\ + \mathbf{Y}^{k^H}\mathcal{B}(\mathbf{U}) \quad (18)$$

where $\mathcal{A}(\mathbf{U}) \triangleq \Omega(\mathbf{FS}[\widehat{\mathbf{P}} \circ (\mathbf{U}\widehat{\mathbf{V}})])$ and $\mathcal{B}(\mathbf{U}) = \mathbf{\Psi U}\widehat{\mathbf{V}}$. Equation (18) is a least-squares problem whose optimal solution can be given by:

$$\left(\mathcal{A}^*\mathcal{A} + \frac{\rho}{2}\mathcal{B}^*\mathcal{B}\right)\mathbf{U}^{k+1} = \mathcal{A}^*(\mathbf{d}) + \frac{\rho}{2}\mathcal{B}^*\left(\mathbf{G}^{k+1} - \frac{\mathbf{Y}^k}{\rho}\right) \quad (19)$$

where $*$ stands for the adjoint operator. Furthermore, assuming the sparsifying transform $\mathbf{\Psi}$ is orthogonal, i.e., $\mathbf{\Psi}^H\mathbf{\Psi} = \mathbf{I}$, (19) can be simplified as:

$$\left(\mathcal{A}^*\mathcal{A} + \frac{\rho}{2}\mathcal{I}\right)\mathbf{U}^{k+1} = \mathcal{A}^*(\mathbf{d}) + \frac{\rho}{2}\mathcal{B}^*\left(\mathbf{G}^k - \frac{\mathbf{Y}^k}{\rho}\right) \quad (20)$$

where $\mathcal{I}(\mathbf{U}) = \mathbf{U}\widehat{\mathbf{V}}\widehat{\mathbf{V}}^H$ is positive definite because $\widehat{\mathbf{V}} \in \mathbb{C}^{L \times N}$ has full row rank. The fact that $\mathcal{A}^*\mathcal{A}$ is also positive definite guarantees (13) to have a unique solution. Equation (13) can be effectively solved using a conjugate gradient algorithm with $\mathbf{U}^k$ as the initialization.

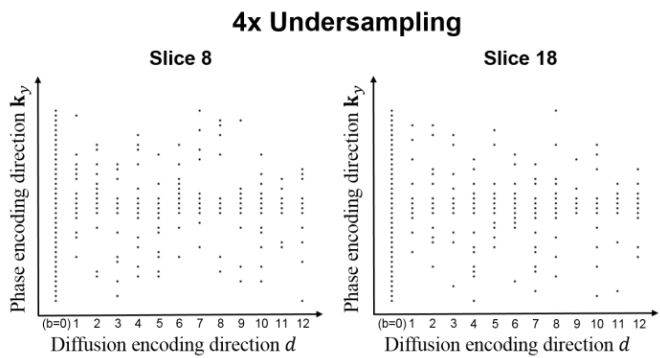

**Fig. 1.** Representative sampling patterns in spatial-diffusion (**k**-$d$) space for two different slices at the acceleration factor of 4.

Note that by applying the ADMM algorithm, the convex optimization problem converges to an optimal solution regardless of the choice of initialization. We initialize $\mathbf{Y}^0$ with a zero matrix out of simplicity, and initialize $\mathbf{U}^0$ with the solution of minimizing only the data consistency term (i.e., $\ell_2$-norm) to decrease the total number of ADMM iterations. Since α soft-thresholds the transform coefficients, we initialize α to be the maximum value in the initial transform domain, and in each subsequent iteration, we reduce α by a fixed amount $c$, i.e., $\alpha^{k+1} = \alpha^k/c$, so that the optimization problem rapidly converges to its global optimal solution. We use $c$=1.55 in this work to assure the fastest convergence. The penalty factor $\rho$ in the $k$th iteration is calculated as $\rho^k = \lambda/\alpha^k$. The stopping criteria is defined as:

$$\|\mathbf{U}^{k+1} - \mathbf{U}^k\|_2 \leq \epsilon \quad \text{and} \quad k > K \quad (21)$$

where $\epsilon$ is the error tolerance between two solutions from consecutive iterations, and $K$ is the maximum number of iterations. In this work, we choose $\epsilon$=10⁻⁹ and $K$=25.

## III. EXPERIMENT DESIGN

### A. Data Acquisition

The proposed method was evaluated both ex vivo and in vivo. For ex vivo experiments, data were acquired from six explanted hearts extracted from heart failure patients. The experiments were approved by the Institutional Review Board (IRB) at Cedars-Sinai Medical Center. The hearts were immersed in saline right after extraction in surgery room, held in place by surrounding towels in a container and collected for imaging 2-4 hours after heart transplant surgery from pathology. Imaging was performed on a 3T Siemens PET/MR Biograph mMR scanner. A diffusion-weighted single spin echo sequence was used for the experiment. The imaging parameters were: TR/TE = 3400/72ms, resolution = 0.9×0.9×2.5mm³, 12 diffusion encoding with b-values of 0 and 1000s/mm², no signal averaging, multi-slice acquisition to achieve whole left ventricle coverage (32, 34 or 38 slices for different datasets) with no gap between slices, total scan time is approximately 3 hours per heart.

In vivo experiments were performed on seven hypertrophic cardiomyopathy patients. The experiments were approved by the IRB at Yonsei University. Free-breathing imaging was performed on a 3T Siemens Prisma scanner with ECG triggering. A second-order motion-compensated (M2) diffusion tensor sequence was used for this experiment, where two pairs of symmetric first-order motion-compensated (M1) diffusion encoding gradients were played out at max gradient strength (80mT/m) to achieve M2 gradient moment nulling [38], followed by a single shot EPI readout. A reduced FOV acquisition was implemented to eliminate signals from other body parts besides the myocardium. The imaging parameters were: TR/TE=430/79ms, resolution = $1.7 \times 1.7 \times 8$mm³, 12 diffusion encoding directions with b-values of 0 and 440s/mm², four signal averages, 10 slices with 8mm slice gap, spatial coverage from basal to apical regions, and total scan time approximately 12 minutes per patient.

### B. k-Space Sampling

We illustrate the sampling pattern we use for reconstruction using joint low-rank and sparsity constraints in Fig. 1. First, for each slice and each diffusion direction, a Gaussian random variable density sampling is applied along the phase encoding direction, considering that the sampling density should be higher at the center of k-space to capture more energy, and to preserve image contrast and the phase information. We additionally acquire four lines at the k-space center for every diffusion direction to ensure that there are a few k-space locations which are densely sampled in (**k**-$d$) space. Second, a different random pattern is applied for each slice and each diffusion direction to increase the incoherence, which especially benefits the compressed sensing component of the proposed method. Third, because the non-diffusion measurement (b=0) is such an important component of image quantification, it is always fully sampled. This provides variable density along the diffusion dimension $d$ and allows simple sensitivity map estimation from the b=0 image.

We use the acceleration factor $R$, defined as the ratio of acquired phase encoding lines and the maximum number of phase encoding lines of the diffusion-weighted k-space signals, to indicate the undersampling level. Because the non-diffusion-weighted signals were fully sampled, the actual acceleration factors evaluated were $R_{\text{true}}$=13 $R/(R+12)$. For convenience, we simply report $R$ in subsequent sections.

### C. Image Reconstruction

For ex vivo datasets, we added zero-mean Gaussian noise to the real and imaginary parts of the raw k-space data to reduce the SNR, from between 25 and 35, to between 12 and 15 (approximating the SNR of in vivo cases). The fully sampled k-space data was retrospectively undersampled using the sampling pattern defined in the previous subsection. Image reconstruction was performed by exploiting low-rankness and group transform sparsity according to (8). Because in vivo imaging was free-breathing, averaging was performed after reconstruction and image registration (when fitting the diffusion tensor), avoiding respiratory blurring. As a result, **X** has $4\times$ as many columns as diffusion directions, and **V** contained contributions from both diffusion weighting and respiratory motion. The rank $L$ was chosen at the "elbow" of the log-scale singular value curve of $\mathbf{P}^* \circ \mathbf{X}$, and was set to be 7 to 8 for ex vivo and 17 to 25 for in vivo. The regularization parameter $\lambda$ was chosen such that the preliminary reconstruction gave the phase map inducing "maximal low-rankness" of the phase-corrected complex coil-combined image as measured by the nuclear norm, i.e., to minimize $\|\mathbf{P}^* \circ \mathbf{X}\|_*$. $\lambda$ and the sensitivity map was the same for all reconstructions. In this work, we chose $\mathbf{\Psi}$ to apply a four-level symlet-4 3D wavelet transform with periodic boundary condition considering that a higher level and 3D transform both contributed to a more sparsified transform domain which enhanced the performance of compressed sensing. Note that any sparsifying transform can be used in place of $\mathbf{\Psi}$ in our framework.

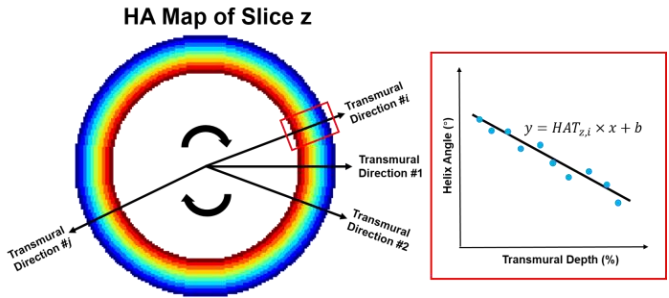

**Fig. 2.** Demonstration for calculation of the helix angle transmurality (HAT) of the zth slice.

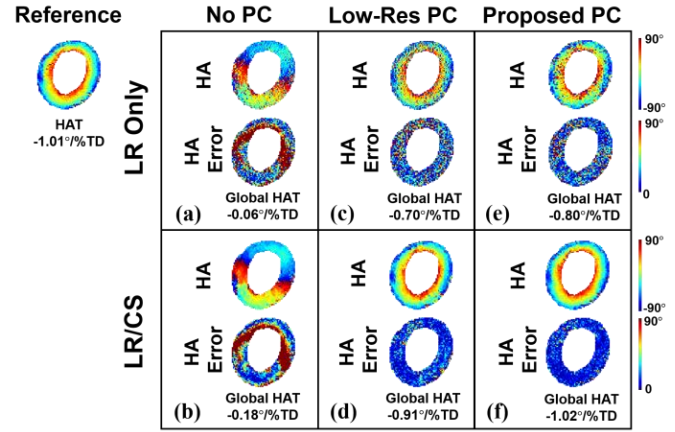

To demonstrate the strength of the proposed method over previous low-rank methods for accelerating DTI, we compared several reconstruction strategies with different phase correction methods: (I) LR Only and LR/CS constrained reconstruction with no phase correction (No PC) inspired from [33]; (II) LR Only and LR/CS constrained reconstruction with low-resolution phase correction (Low-Res PC) inspired from [27], where half of the sampled lines fully sampled the center of k-space for phase estimation and the remainder of the sampled lines were collected using lattice sampling as in [27]; and (III) LR Only and LR/CS constrained reconstruction with the proposed phase correction. We also demonstrated the advantage of using the joint constraints (LR/CS) over using only group sparsity constraint (CS Only) or only low-rank constraint (LR Only) with the proposed phase correction procedure. The image reconstruction was performed on a Linux workstation with a 2.70GHz dual 12-core Intel Xeon processor equipped with 256 GB RAM and running MATLAB R2015b. Each reconstruction took 15-20 minutes.

*D. Image Analysis*

**Fig. 3.** Comparison between phase correction strategies for ex vivo. Reconstructed HA maps at $R=2$ with corresponding voxel-wise HA error **(a-b)** without phase correction using LR Only and LR/CS, **(c-d)** with low-resolution phase correction using LR Only and LR/CS, **(e-f)** with the proposed phase correction using LR Only and LR/CS.

The myocardium of the left ventricle was chosen to be the region of interest (ROI). The diffusion tensor, along with some conventional derived metrics showing the fiber orientation and organization such as HA, helix angle transmurality (HAT) as well as MD was calculated and compared between the reference and the reconstructed images at varying $R$. HAT is a measurement characterizing the transmural continuum of HA from endocardium to epicardium, denoted as the slope of HA over transmural depth (TD) across the myocardium. It was calculated by radially sampling the HA along 25 transmural directions and fitting the line using linear regression (Fig. 2).

*E. Quantitative Analysis*

We calculated the global HAT (i.e., the average HAT of the whole left ventricle) and the global MD (i.e., the average MD value in the ROI of the whole left ventricle) to derive the normalized bias

$$\beta = |(h_{\text{rec}} - h_{\text{ref}})/h_{\text{ref}}| \tag{23}$$

where $h_{\text{ref}}$ denotes the global HAT (or global MD) of the reference datasets and $h_{\text{rec}}$ denotes the global HAT (or global MD) of the reconstructed datasets, and intra-class correlation coefficients $r$ which measures the degree of absolute agreement between reference and reconstructed global HAT and global MD. ICC was calculated using IBM SPSS Statistics with a two-way mixed model and a confidence level of 95%. The criteria for ICC measurements is as follows [39]:

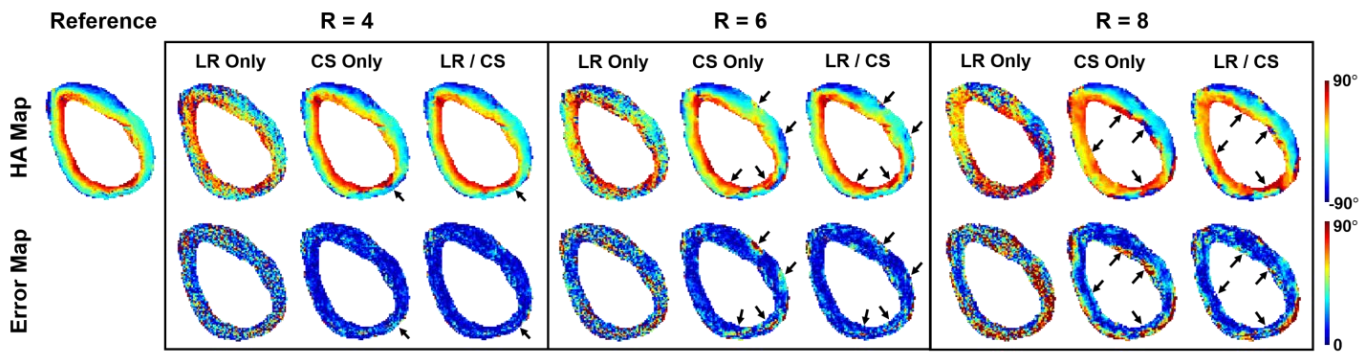

**Fig. 4.** Reconstructed ex vivo HA maps and the corresponding voxelwise error maps at acceleration factors of $R$=4.0, 6.0 and 8.0 using LR Only, CS Only and LR/CS. The black arrows point to the regions where distinguishable between LR/CS and CS Only.

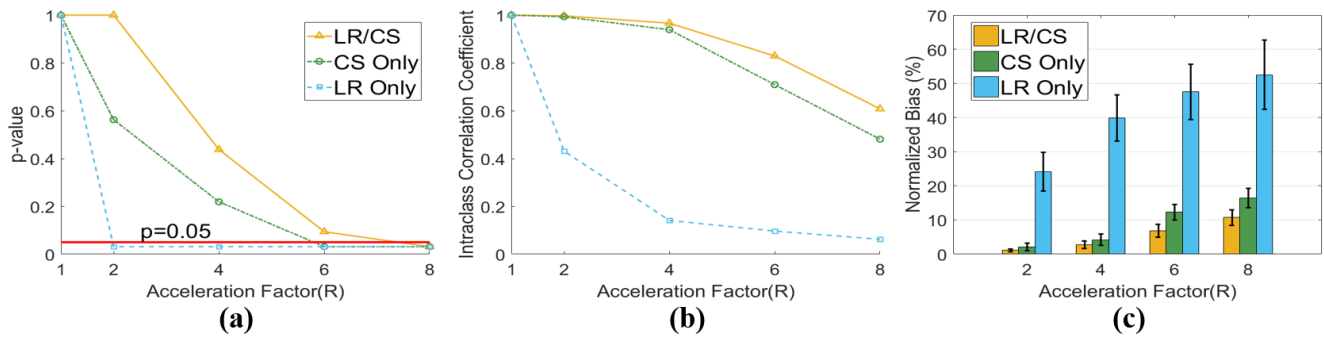

**Fig. 5.** Statistical analysis of global HAT based on all six ex vivo datasets using LR Only, CS Only and LR/CS. **(a)** p-value analysis between reference and reconstructed global HAT at acceleration factors of $R=2.0, 4.0, 6.0$ and $8.0$. The significance level ($P=0.05$) is labeled in red. **(b)** The intra-class correlation coefficient analysis between reference and reconstructed global HAT. **(c)** Normalized bias between reference and reconstructed global HAT.

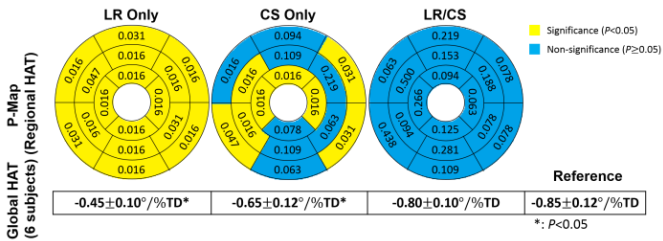

**Fig. 6.** Ex vivo p-maps of regional HAT using LR Only, CS Only and LR/CS at 6× acceleration based on 16 AHA segments of the left ventricle and corresponding global HAT statistics across 6 subjects.

- Less than 0.40—Poor
- Between 0.40 and 0.59—Fair
- Between 0.60 and 0.74—Good
- Between 0.75 and 1.00—Excellent

In addition, a p-value analysis for all ex vivo datasets were also performed using a Wilcoxon signed rank test between reference and reconstructed global HAT and global MD at each $R$ to determine the highest potential acceleration factors for LR Only, CS Only and LR/CS. The significance level was set to be the conventional level $P=0.05$. For each method, those $R$ corresponding to p-values above the significance level were considered feasible acceleration factors.

Quantitative regional statistical analysis was performed both ex vivo and in vivo for each reconstruction method. We first calculated regional HAT and MD values of 16 AHA segments (6 basal segments, 6 mid segments, 4 apical segments) for all subjects, and then performed p-value analysis for each segment, generating a p-map that indicated the ability of each reconstruction method to preserve regional HAT and MD measurements without significance from the reference. The significance level for regional analysis was also $P=0.05$.

## IV. RESULTS

### A. Ex vivo Cardiac Diffusion Tensor Imaging

*1) Comparisons against different phase correction strategies in existing low-rank methods:* Fig. 3 demonstrates the HA maps and global HAT measurements of using different phase correction methods at 2× acceleration with an illustrative example of one subject. Without phase correction, the reconstructed HA maps are the least accurate, with the helical fiber structure completely altered and the transmural continuum disrupted (Global HAT=-0.06°/%TD for LR Only vs Global HAT=-1.01°/%TD for reference: 94% error, and HAT=-0.18°/%TD for LR/CS vs Global HAT=-1.01°/%TD for reference: 82% error). The application of a low-resolution phase map results in enhanced quality of the HA maps with partly restored transmural change (Global HAT=-0.70°/%TD for LR Only vs Global HAT=-1.01°/%TD for reference: 31% error, Global HAT=-0.91°/%TD for LR/CS vs Global HAT=-1.01°/%TD for reference: 10% error). The proposed phase correction strategy shows advantages over low-resolution phase correction, yielding the most accurate reconstructed HA maps and the transmural continuum (Global HAT=-0.80°/%TD for LR Only vs Global HAT=-1.01°/%TD for reference: 21% error, Global HAT=-1.02 °/% TD for LR/CS vs Global HAT=-1.01°/%TD for reference: 1% error).

*2) Helix angle and helix angle transmurality:* At a moderate acceleration factor ($R=4.0$), both LR/CS and CS Only preserves the helical angle features and the transmural continuum across the myocardium (Fig. 4). At aggressive acceleration factors ($R \geq 6.0$), however, LR/CS preserves HA features with less reconstruction errors at the inferior wall and the lateral wall, compared to CS Only. LR Only, on the other hand, results in noise and severe image artifacts consistently.

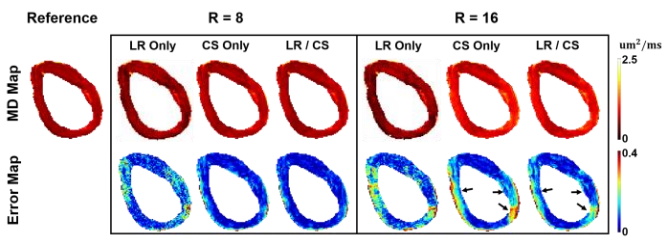

**Fig. 7.** Reconstructed ex vivo MD maps and the corresponding voxelwise error maps at $R=8.0$ and $16.0$. The black arrows point to the regions where distinguishable between LR/CS and CS Only.

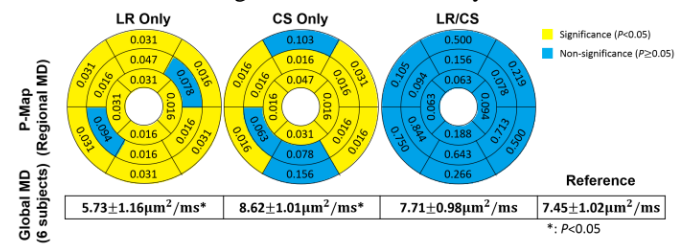

**Fig. 8.** Ex vivo p-maps of regional MD using LR Only, CS Only and LR/CS at 16× acceleration based on 16 AHA segments of the left ventricle and corresponding global MD statistics across 6 subjects.

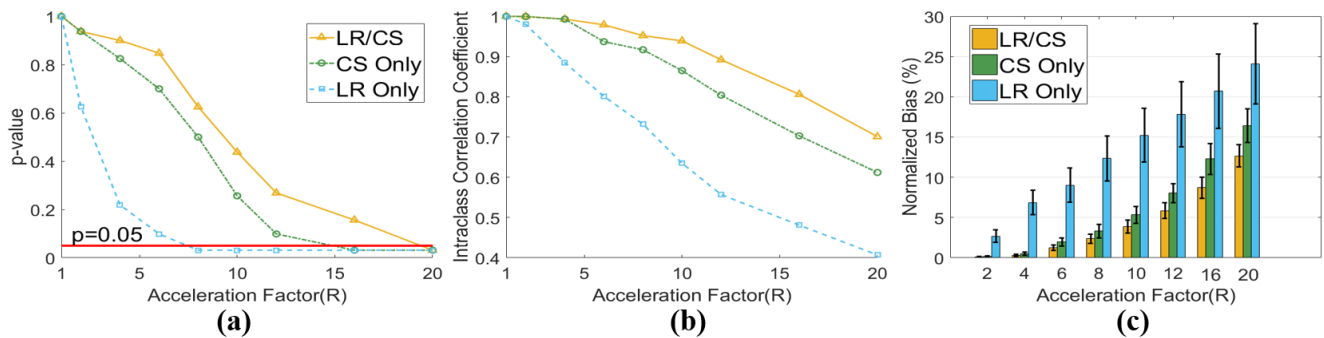

**Fig. 9.** Statistical analysis of global MD based on all six ex vivo datasets using LR Only, CS Only and LR/CS. **(a)** p-value analysis between reference and reconstructed global MD at acceleration factors of $R$=2.0, 4.0, 6.0, 8.0, 10.0, 12.0, 16.0 and 20.0. The significance level ($P$=0.05) is labeled in red. **(b)** The intra-class correlation coefficient analysis between reference and reconstructed global MD. **(c)** Normalized bias between reference and reconstructed global MD.

For p-value analysis of global HAT, CS Only and LR/CS are able to yield reconstructed global HATs without significant difference ($P$>0.05) from the reference at $R$=4.0 and $R$=6.0, respectively (Fig. 5a). Thus, the maximum feasible acceleration factors for CS Only and LR/CS are $R$=4.0 and $R$=6.0. LR Only, however, yields significantly different ($P$<0.05) global HAT even at the most moderate $R$=2.0.

For ICC analysis of global HAT, LR Only consistently gives "Fair" or "Poor" ICC (<0.50). CS Only results in "Excellent" ICC (0.99, 0.94) at moderate acceleration factors $R$=2.0 and 4.0, "Good" ICC (0.71) at more aggressive $R$=6.0 and "Fair" ICC (0.48) at $R$=8.0, while LR/CS consistently yields higher ICC, with "Excellent" ICC (0.99, 0.97, 0.83) for $R \leq$6.0 and "Good" ICC (0.61) at $R$=8.0 (Fig. 5b).

For bias analysis of global HAT, LR Only yields significantly larger bias (>20%, $P$<0.05) at all $R$. LR/CS yields less bias than CS Only (Fig. 5c). Specifically, both methods yield small bias (<5%) at $R \leq$4.0, showing no significant differences ($P$=0.69 at $R$=2.0, $P$=0.16 at $R$=4.0) between each other. At $R$=6.0, LR/CS yields significantly lower bias than CS Only (6.8%$\pm$1.9% vs 12.3%$\pm$2.3% $P$=0.03).

At $R$=6.0, LR Only yields significance in global HAT (all 6

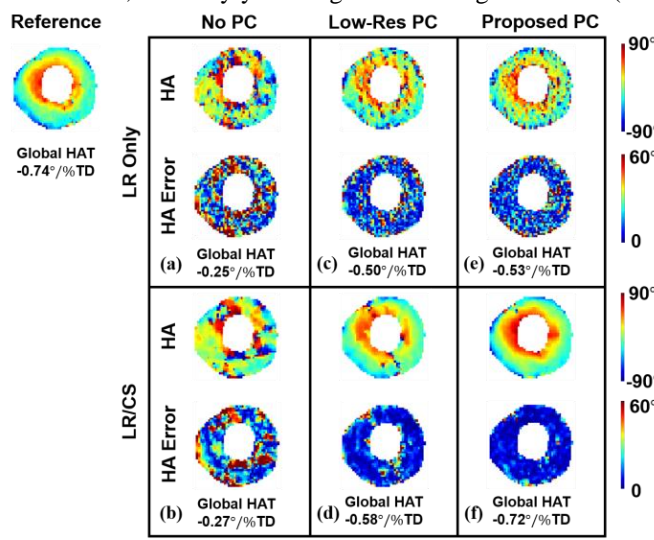

**Fig. 10.** Comparison between phase correction strategies for in vivo. Reconstructed HA maps at $R$=2 with corresponding voxel-wise HA error **(a-b)** without phase correction using LR Only and LR/CS, **(c-d)** with low-resolution phase correction using LR Only and LR/CS, **(e-f)** with the proposed phase correction using LR Only and LR/CS.

subjects: -0.45$\pm$0.10°/%TD vs -0.85$\pm$0.11°/%TD, $P$=0.03), and exhibits significant differences ($P$<0.05) in regional HAT measurements for all 16 AHA segments. CS Only also yields significance in global HAT (all 6 subjects: -0.65$\pm$0.12°/%TD vs -0.85$\pm$0.11°/%TD, $P$=0.03), but preserves regional HAT measurements without significance ($P$>0.05) at 8 segments out of 16. LR/CS provides the best performance, yielding global HAT without significance (all 6 subjects: -0.80$\pm$0.10°/%TD vs -0.85$\pm$0.11°/%TD, $P$=0.06), and preserving regional HAT measurements without significance ($P$>0.05) at all 16 segments (Fig. 6).

*3) Mean diffusivity:* MD was robust to acceleration factors, with bias decreasing as $R$ increases compared to HAT (Fig. 5c and Fig. 9c). As a result, we explore higher acceleration factors (up to 20.0). To demonstrate robustness, Fig. 7 shows the MD maps at $R$=8.0 and $R$=16.0. At $R$=8.0, CS Only and LR/CS gives almost identical results with small bias (3% to 4%) across the myocardium. At $R$=16.0, LR/CS yields less errors around the septal and lateral wall. LR Only produces significant noise and results in large errors across the myocardium compared to the other two methods.

For the p-value analysis of the global MD, the maximum feasible acceleration factors for LR Only, CS Only and LR/CS are $R$=6.0, $R$=12.0 and $R$=16.0, respectively, considering no significant differences ($P$>0.05) (Fig. 9a).

LR/CS yields higher or equal ICC than CS Only all the way up to $R$=20.0 (Fig. 9b). Both methods are able to yield "Excellent" ICC ($\geq$ 0.80) at $R \leq$ 12.0. The maximal acceleration factor that allows for LR/CS to give an "Excellent" ICC (0.81) is 16.0, higher than CS Only (ICC=0.80 at $R$=12.0). LR Only yields lower ICC over all compared to the other two methods and gives "Excellent" ICC only at $R \leq$6.0.

LR/CS yields less bias than CS Only at all the acceleration factors (Fig. 9c). Both methods yield less than 5% bias at $R \leq$ 10.0, beyond which the bias given by LR/CS is significantly lower than by CS Only (5.8% $\pm$ 0.9% versus 8.1% $\pm$ 1.2%, $P$ =0.03 at $R$ =12.0, 8.7% $\pm$ 1.3% versus 12.3%$\pm$1.9%, $P$=0.03 at $R$=16.0). LR Only consistently yields significantly higher bias than CS Only and LR/CS at all $R$.

At $R$=16.0, LR Only yields significance in global MD (all 6 subjects: 5.73$\pm$1.16um$^2$/ms vs 7.45$\pm$1.02um$^2$/ms, $P$=0.03), and produces significant differences ($P$<0.05) in regional MD measurements at 14 segments out of 16. CS Only also yields significance in global MD (all 6 subjects: 8.62$\pm$1.01um$^2$/ms vs 7.45±1.02um$^2$/ms, $P$=0.03), and yields significance ($P$<0.05) in regional MD measurements at 12 segments out of 16. LR/CS performs the best, showing no significance in global MD (all 6 subjects: 7.71±0.98um$^2$/ms vs 7.45±1.02um$^2$/ms, $P$=0.15), and preserving regional MD measurements at all 16 segments without significance ($P$>0.05) (Fig. 8).

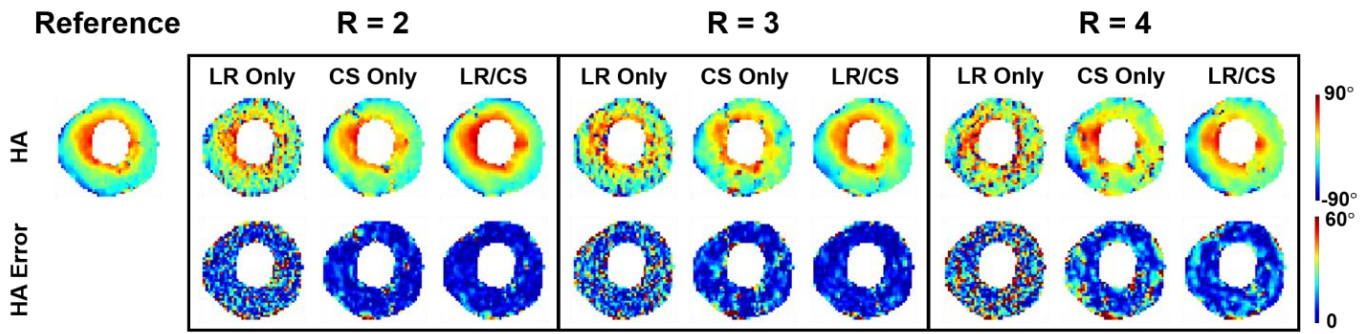

**Fig. 11.** Reconstructed in vivo HA maps and the corresponding voxelwise error maps at acceleration factors of $R$=2.0, 3.0 and 4.0 using LR Only, CS Only and LR/CS.

### B. In vivo Cardiac Diffusion Tensor Imaging

*1) Comparisons against different phase correction strategies in existing low-rank methods:* Fig. 10 compares different phase correction methods associated with low-rank modeling at 2× acceleration with an illustrative example of one subject. Without phase correction, HA maps exhibit large errors and disrupted transmural continuum with and without CS (Global HAT=-0.18°/%TD for LR Only vs Global HAT=-0.74°/%TD for reference: 76% error, and Global HAT=-0.27°/%TD for LR/CS vs Global HAT=-0.74°/%TD for reference: 64% error). Low-resolution phase correction improves reconstructed HA features and the transmural change (Global HAT=-0.50°/%TD for LR Only vs Global HAT=-0.74°/%TD for reference: 32% error, and Global HAT=-0.58°/%TD for LR/CS vs Global HAT=-0.74°/%TD for reference: 22% error). The proposed phase correction strategy preserves HA features with the most accuracy, resulting in well-preserved transmural continuum with the addition of CS (Global HAT=-0.53°/%TD for LR Only vs Global HAT=-0.74°/%TD for reference: 28% error, and Global HAT=-0.72°/%TD for LR/CS vs Global HAT=-0.74°/%TD for reference: 3% error).

*2) Helix angle and helix angle transmurality:* HA maps from LR Only, CS Only and LR/CS reconstructions at $R$=2.0, 3.0 and 4.0 are compared (Fig. 11). LR Only results in noise and image artifacts consistently at all acceleration factors. CS Only suppresses noise, offering better reconstruction quality and representation of HA features, but still contains artifacts. LR/CS further removes the residual artifacts and yields the best reconstruction quality with the least amount of errors across the myocardium compared to CS Only, thus producing the most accurate HA map and transmural changes across the myocardium.

Analysis of global HAT measurements reveals that for in vivo cases, the highest feasible acceleration factor providing no significant difference from the reference ($P$>0.05) is $R$=3.0 for LR/CS, which is higher than CS Only ($R$=2.0) and LR Only ($R$<2.0). LR/CS provides "Excellent" ICC up to $R$=3.0 (0.89 at $R$=2.0, 0.83 at $R$=3.0), while CS Only provides "Excellent" ICC (0.81) only at $R$=2.0 and LR Only yields "Poor" ICC (<0.40). LR/CS produces significantly lower bias than CS Only at feasible $R$ (4.3%±2.1% versus 10.1%±3.3%, $P$=0.01 at $R$=2.0, 10.3%±4.1% versus 27.9%±10.0%, $P$=0.01 at $R$=3.0), while LR Only consistently yields significantly higher bias (Fig 12).

At $R$=3.0, LR Only yields significance in global HAT measurements (all 7 subjects: -0.27±0.13 °/% TD vs -0.62±0.12 °/% TD, $P$=0.01), and exhibits significant differences ($P$<0.05) in regional HAT measurements at all 16 segments. CS Only also yields significance in global HAT measurements (all 7 subjects: -0.44±0.09 °/% TD vs -0.62±0.12 °/% TD, $P$=0.01), and produces significant differences ($P$<0.05) in regional HAT measurements at 9 segments out of 16. LR/CS yields no significance in global HAT measurements (all 7 subjects: -0.57±0.11°/%TD vs -0.62±0.12°/%TD, $P$=0.08), and preserves regional HAT

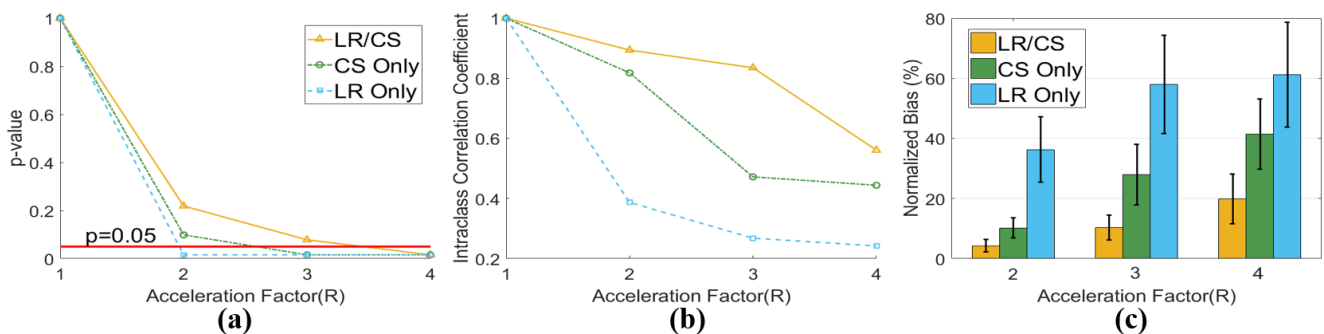

**Fig. 12.** Statistical analysis of global HAT based on all seven in vivo datasets using LR Only, CS Only and LR/CS. **(a)** p-value analysis between reference and reconstructed global HAT at acceleration factors of $R$=2.0, 3.0 and 4.0. The significance level ($P$=0.05) is labeled in red. **(b)** The intra-class correlation coefficient analysis between reference and reconstructed global HAT. **(c)** Normalized bias between reference and reconstructed global HAT.

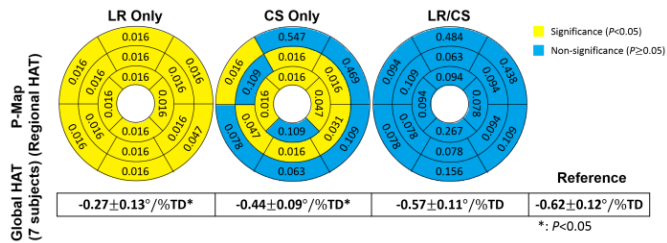

**Fig. 13.** In vivo p-maps of regional HAT using LR Only, CS Only and LR/CS at 3× acceleration based on 16 AHA segments of the left ventricle and corresponding global HAT statistics across 7 subjects.

measurements without significance ($P>0.05$) at all 16 segments (Fig. 13).

*3) Mean diffusivity:* MD maps from LR Only, CS Only and LR/CS reconstructions at $R=2.0$, 3.0 and 4.0 are compared (Fig. 14). LR Only induces corrupted MD maps at all $R$ due to noise and artifacts. CS Only provides regularization and smooths the maps, but shows significant increase in MD as $R$ goes up, yielding large errors (almost >20% across the whole myocardium) at $R=4.0$. LR/CS further improves the reconstruction, substantially reducing the errors (<5% across the myocardium at $R \leq 3.0$, around 10% at $R=4.0$) and preserving more accurate MD features.

Analysis on global MD measurements reveal that, for in vivo, LR/CS is able to reach a higher acceleration factor of $R=4.0$ without significant difference from the reference ($P>0.05$), compared to CS Only ($R=2.0$) and LR Only ($R<2.0$). LR/CS provides "Excellent" ICC (>0.75) all the way up to $R=4.0$, while CS Only does only at $R=2.0$. LR Only yields "Poor" ICC consistently at all $R$. LR/CS yields significantly lower bias than CS Only at feasible acceleration factors (0.6%$\pm$0.4% versus 3.1% $\pm$ 1.5%, $P=0.03$ at $R=2.0$, 2.5% $\pm$ 1.3% versus 8.6% $\pm$ 2.4%, $P=0.01$ at $R=3.0$, 7.0% $\pm$ 1.6% versus 13.5%$\pm$3.0%, $P=0.01$ at $R=4.0$). LR Only yields significantly higher ($P<0.05$) bias than CS Only and LR/CS (Fig. 15).

At $R=4.0$, LR Only yields significance in global MD (all 7 subjects: 1.83$\pm$0.21mm$^2$/s vs 2.31$\pm$0.14mm$^2$/s, $P=0.01$), and produces significance ($P<0.05$) in regional MD at 14 segments out of 16. CS Only also yields significance in global MD (all 7 subjects: 2.80$\pm$0.22mm$^2$/s vs 2.31$\pm$0.14mm$^2$/s, $P=0.01$), and produces significance ($P<0.05$) in regional MD at 14 segments out of 16. LR/CS provides the best performance, yielding no significance in global MD (all 7 subjects: 2.37$\pm$0.18mm$^2$/s vs 2.31 $\pm$ 0.14mm$^2$/s, $P=0.15$), and preserving regional MD without significance ($P>0.05$) at all 16 segments (Fig. 16).

## V. DISCUSSIONS

We proposed a novel method to accelerate CDTI in a framework jointly combining low-rank and spatial sparsity constraints. We performed the experiment on six explanted diseased ex vivo human heart failure cadavers and seven hypertrophic cardiomyopathy patients, and introduced a population-based statistical analysis on global CDTI measurements which included the p-value analysis, bias analysis, and intra-class correlation coefficient analysis which were used to determine the highest acceleration factor that preserves heart failure features, the statistical deviation from the reference, and the absolute agreement with the reference. We also performed quantitative analysis on regional CDTI measurements to evaluate the proposed method against employing either single constraint alone.

CDTI has limited applications in clinical environment due to one major challenge – the prolonged acquisition time. A distributed compressed sensing technique has been applied to accelerate ex vivo CDTI by leveraging a group sparsity penalty [22]. This method allows for up to 4× acceleration but beyond which image quality and reconstruction accuracy begin to decrease. We exploited the correlated diffusion behaviors across directions and incorporated low-rank modeling with phase correction procedure to the compressed sensing framework, and demonstrated the strength of the proposed method (LR/CS) over using compressed sensing only (CS Only) and low-rank only (LR Only). Both ex vivo and in vivo results have shown that the proposed method outperformed the use of either single constraint alone, removing image artifacts and restoring image features of both HA and MD while being able to achieve higher acceleration factors, showing more substantial absolute agreement with the reference and less bias from the reference. In addition, LR/CS was able to preserve accurate regional HAT and MD measurements without significance from the reference at all 16 AHA segments at high acceleration factors.

In this study, global HAT limited the acceleration to $R=6.0$ for ex vivo and $R=3.0$ for in vivo while global MD exhibited robustness to acceleration. This was due to the uniformness of MD across the myocardium. The image resolution decreased as the acceleration factor increased, causing the sinc-shaped point spread function to stretch out, allowing signals outside the myocardium to alias into the myocardium along phase encoding direction, which started to alter the diffusion property of the myocardium from epicardial edges of septal and lateral walls where MD errors occurred and gradually entered the myocardium. MD of anterior and inferior walls did not alter much, as the uniform signals from such myocardial regions remained inside the myocardium.

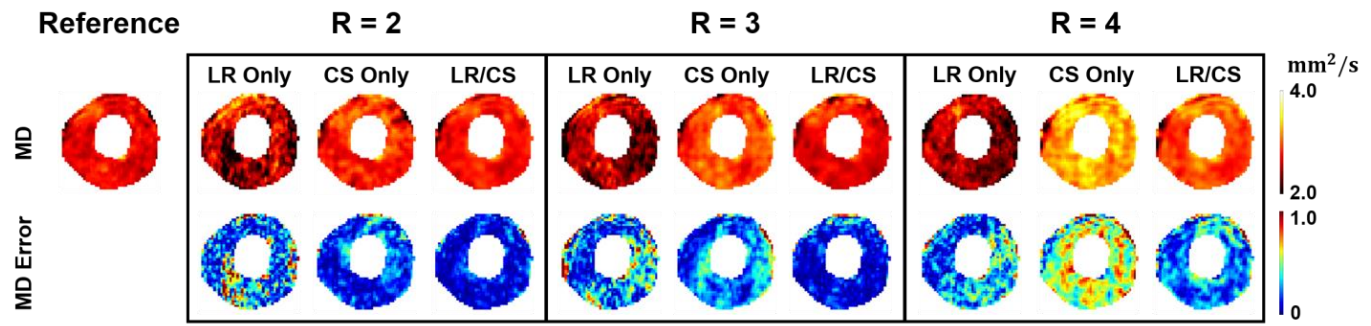

**Fig. 14.** Reconstructed in vivo MD maps and the corresponding voxelwise error maps at acceleration factors of $R=2.0$, 3.0 and 4.0 using LR Only, CS Only and LR/CS.

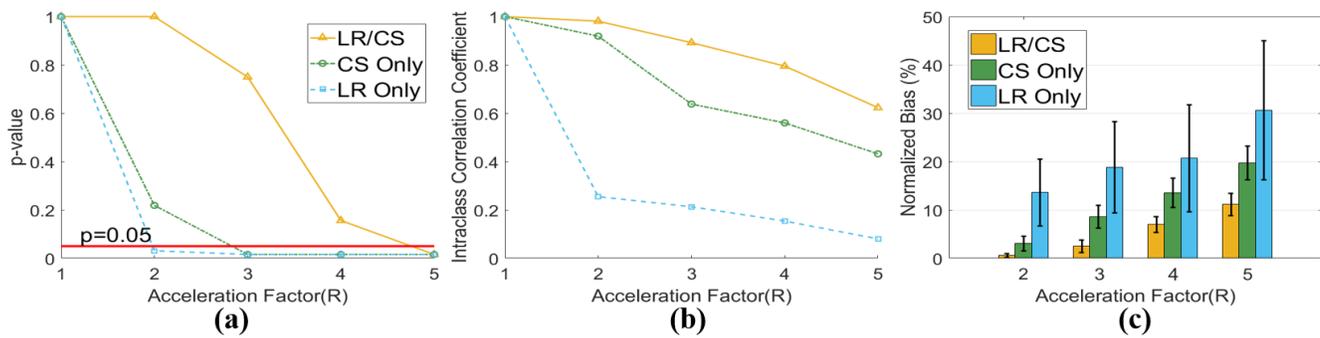

**Fig. 15.** Statistical analysis of global MD based on all seven in vivo datasets using LR Only, CS Only and LR/CS. **(a)** p-value analysis between reference and reconstructed global MD at acceleration factors of $R$=2.0, 3.0, 4.0 and 5.0. The significance level ($P$=0.05) is labeled in red. **(b)** The intra-class correlation coefficient analysis between reference and reconstructed global MD. **(c)** Normalized bias between reference and reconstructed global MD.

In vivo datasets demonstrated less acceleration than ex vivo datasets potentially due to differences in spatial resolution and the presence of respiratory motion. Ex vivo data were acquired with higher in-plane spatial resolution ($0.9 \times 0.9 mm^2$ vs $1.7 \times 1.7 mm^2$), substantially more phase encoding lines (64 lines vs 192 lines), and substantially longer scan time (3 hours per heart vs 12 minutes per patient) than in vivo data. Taken together, the degree of acceleration for in vivo datasets were limited due to overall SNR despite our efforts to artificially increase noise in ex vivo datasets to match in vivo SNR. Furthermore, the presence of respiratory motion during in vivo imaging weakened the low-rankness of the in vivo images by potentially reducing the correlation between images and increasing the degrees of freedom compared to motion-free cases, reducing the overall potential for acceleration.

We demonstrated that incorporating low-rank modeling to a compressed sensing framework was beneficial. Low-rank modeling and compressed sensing are complementary to each other. It is worth mentioning that ex vivo datasets were all motion-free, which revealed that the low-rank constraint was able to characterize the signal correlation induced by diffusion process, resulting in fewer degrees of freedom which facilitated undersampling. In vivo datasets, however, experienced both the diffusion process and respiratory motion. We demonstrated that the low-rank modeling with the proposed phase correction method was also able to characterize the signal correlation induced by the joint effect of diffusion and respiration. The proposed method has the potential to be applied to in vivo clinical research, increasing image resolution, reducing temporal footprints, and achieving a reduction of scan time via a prospective implementation.

Another important aspect worth discussing is the phase correction procedure. Previous works on accelerating DTI have either combined an implicit low-rank constraint and compressed sensing without phase correction [33], or employed a single implicit low-rank constraint, with a low-resolution phase map pre-estimated from central k-space data [27]. In this work, we performed a preliminary reconstruction (CS Only) using all the acquired data to estimate the phase map. Without phase correction, the drastic phase inconsistency across diffusion directions were not addressed, resulting in uncorrelated complex diffusion weighted signals that weakened low-rankness, therefore inducing significant reconstruction errors and destroying the helical fiber structure. The low-resolution phase correction partly restored the signal coherence, reducing errors and recovering the transmural continuum across the myocardium. The proposed phase correction took advantage of all the acquired data, yielded less error in the ROI and improved transmural change across the myocardium. This indicated that the proposed phase correction had stronger ability to address the phase inconsistency and restore signal coherence, thus enhancing low-rankness.

For the low-rank constraint in the proposed method, we implemented an explicit rank constraint where a pre-estimated low-dimensional diffusion subspace $V$ was required. Conventional methods to estimate $V$ require auxiliary training data as described in [23, 24]. In this work, the magnitude part of the preliminary reconstruction (CS Only) was used to estimate $V$, which eliminated phase inconsistencies. In addition, all the acquired data was used to estimate $V$, resulting in a more precise subspace. The explicit low-rank constraint results in a simple computational problem. One alternative is to use implicit rank constraint where a nuclear norm or a Schatten p-norm is regularized to constrain the low-rankness [29].

The proposed method required selection of a rank $L$ and a regularization parameter $\lambda$. For each dataset, we chose $L$ independently based on the rank analysis of the reference (i.e. the full-sampled complex coil-combined image). In practice, it may be preferable to automatically set $L$ according to criteria such as Stein's unbiased risk estimation (SURE) [40] or the Akaike information criterion [41]. Similarly, there are criteria to estimate $\lambda$ as well, such as generalized cross-validation method based on (SURE) [40] or the L-curve method [42].

## VI. CONCLUSIONS

In this work, we propose a novel reconstruction framework to accelerate cardiac diffusion tensor imaging (CDTI) by combining phase-corrected joint low-rank and spatial sparsity constraints. Through experiments on six explanted human heart failure cadavers and seven hypertrophic cardiomyopathy patients, we demonstrate that the combination of low-rank and sparsity constraints results in higher acceleration of both ex vivo and in vivo CDTI, improved preservation of helix angle and mean diffusivity features, global and regional helix angle transmurality and mean diffusivity measurements, higher reconstruction accuracy with significantly lower bias, and more substantial absolute agreement with the fully sampled datasets compared to using either single constraint alone. Prospective implementation of the proposed method has the potential to increase image resolution, reduce temporal footprints and/or reduce the scan time of in vivo clinical CDTI studies.

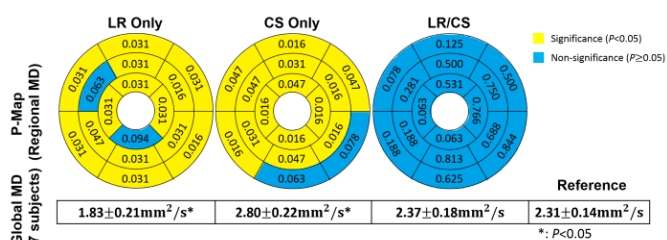

**Fig. 16.** In vivo p-maps of regional MD using LR Only, CS Only and LR/CS at 4× acceleration based on 16 AHA segments of the left ventricle and corresponding global MD statistics across 7 subjects.